\documentclass[letterpaper,aps,floatfix,twocolumn]{revtex4}
\usepackage{enumitem}
\usepackage{graphicx}
\usepackage{xcolor}
\usepackage{blindtext}
\usepackage{grffile}
\usepackage{epsfig}
\usepackage{amsmath}
\usepackage{amssymb}
\usepackage{hyperref}
\usepackage{alltt}
\usepackage{setspace}
\usepackage[english]{babel}
\usepackage{euscript}
\usepackage{tikz}
\usepackage[T1]{fontenc}
\usepackage[latin9]{inputenc}
\usepackage{yfonts}

\newcommand{\old}[1]{{}}

\begin{document}

\title{Verification and Validation of Log-Periodic Power Law Models}

\author{Jarret Petrillo}
\email{jarret.petrillo@stonybrook.edu}
\affiliation{Stony Brook University, Stony Brook, NY 11794 and GlimmAnalytics LLC}

\date{\today}

\begin{abstract}
	We propose and implement a nonlinear Verification and Validation (V\&V) methodology to test two fitting procedures for the log-periodic power law model (LPPL), a model that has diverse applications across data analysis, but known estimation issues.
	Prior studies have focused on ex-post analyses of rare events: Earthquakes, glacial break-off events, and financial crashes.
	Or, on non-dynamical simulations such as additive noise or resampling.
	Our results reject an estimation scheme that pre-conditions observed data by fitting and removing an exponential trend.
	We validate a subordinated algorithm, and confirm that it passes Feigenbaum's criticism, which articulates a broad hurdle for ex-post statistical learning from rare events.
\end{abstract}

\maketitle

\section{Introduction}
Log-periodic power law (LPPL) models second-order, super-exponential behavior near a phase transition.
It has been used to study behavior of the Earth's crust in the vicinity of earthquakes \cite{SalSamSor96, SorSam95, SorMeaWhe20};
glacial break-off events \cite{FaiFunVag16}; real estate prices \cite{ZhoSor04}; and equity bubbles \cite{JohLedSor98, BotMei03, MarketCrash, JohSor99}.

LPPL models are poorly conditioned \cite{BreChaPei13, FilSor13} and the vast majority of studies are ex-post analyses \cite{ZhoSor04, SalSamSor96, SorSam95, SorMeaWhe20, MarketCrash, JohLedSor98, JohSor99}.
Two notable exceptions are \cite{FaiFunVag16}, which presents the results of a real-time monitoring for a glacial break-off event, and
\cite{LalPot99}, which presents the negative result of a trading decision using an LPPL model fit.

Our contribution is the first LPPL Verification and Validation (V\&V) methodology to use nonlinear simulational data.
Simulational studies hitherto have been restricted to white or auto-regressive noise added to historical data \cite{FaiFunVag16, BreChaPei13}, GARCH(1,1) simulations \cite{JohSorLed99} or bootstrap resampling of historical data \cite{BowOui98}.
Our V\&V methodology is used to study two LPPL parameter estimation algorithms.

The paper is organized as follows: section two presents the theory (\ref{sec:lppl-theory}) and practical advice (\ref{sec:practical}) for fitting the LPPL model, section three develops the ABCDE model (\ref{sec:ABCDE}) and its application to non-linear phase transition simulations (\ref{sec:phase}), section four presents the statistical methodology (\ref{sec:methodology}), section five contains the results and further discussion (\ref{sec:results}), and section six concludes (\ref{sec:conclusion}).

\section{Log-Periodic Power Law Model}
\subsection{Theory}\label{sec:lppl-theory}

If $s(t)$ is a time series, the log-periodic power law model takes the form
\begin{equation}
	\label{eq:lppl}
	s(t) = A + B(t_c-t)^m + C(t_c-t)^m \cos(\omega \ln(t_c -t) - \psi) 
\end{equation}
where $t_c$ is a critical transition, $\psi$ is a phase parameter, $\omega$
is frequency, $A,B,C$ are constants, and $m$ is the exponential decay.

The model has many interpretations.  If $s(t)= \log p(t)$, a log asset price, then the LPPL model can be successfully motivated as the solution to a Wiener price process with jumps and heterogeneous market participants \cite{FilSor13, JohLedSor98}.
If $s(t)$ is an abstract dynamical process, then (\ref{eq:lppl}) describes a discrete scale invariant process \cite{SorSam95, SalSamSor96} nearing an exponential phase transition \cite{VanBov98}.

The LPPL model is also understood as a second-order expansion about the standard power law \cite{JohSor99,ZhoSor04}
\begin{equation}
	s(t) = A+B(t_c-t)^m
\end{equation}
The additional terms in the LPPL model, $C, \omega$, and $\psi$, describe additive oscillations in log-periodic frequency, phase, and amplitude, respectively $\omega, \psi$, and $C$.
As $ C\rightarrow 0$ the LPPL is functionally a standard power law.

\subsection{Parameter Estimation}
\label{sec:practical}

Observational data $\{ x_1, \dots, x_n \}$ lead to parameter estimates $\hat{t}_c$, $\hat{\omega}$, $\hat{m}$, $\hat{A}$, $\hat{B}$, $\hat{C}$, and $\hat{\psi}$.
Fitting the seven parameters presents a challenge when framed as a non-linear least-squares minimization problem.
The seven-parameter optimization space has been found to contain both extremely sensitive (stiff) and insensitive parameter axes \cite[Fig.~2]{FilSor13}.
A stability of parameter estimates has been achieived by using median estimates over various sample windows \cite{BreChaPei13}, and by designing estimation procedures that reduce the dimensions.

\subsubsection{Subordinated Estimation Algorithm}

The subordinated algorithm, \textit{subordinates} the linear variables to the nonlinear ones.
A discovery \cite{FilSor13} allows the linear parameters, $A,B$, $C$, and $\psi$, to be found as the solutions to a linear system dependent on the remaining nonlinear parameters.
The authors \cite{FilSor13} suggest a third tier of subordination, whereby $\hat{t}_c$ is first estimated as the value that would minimize a least-squares estimation procedure varying $m$ and $\omega$.
This algorithm can also be repeated over various sample windows.
The data $\{ x_{i_1}, \dots, x_{i_n} \}$ is subindexed by the subsample $i$, where $i_n$ is also understood to vary with the sample, i.e. allow various length periods.\\
\\
\textbf{Subordinated Algorithm:}
\begin{equation*}
	\{ x_{i_1} , \dots, x_{i_n} \} \rightarrow \{ \hat{t}_{c_i} \} \rightarrow \{ \hat{\omega}_i,\hat{m}_i \} \rightarrow \{\hat{A}_i,\hat{B}_i,\hat{C}_i, \hat{\psi}_i\}
\end{equation*}
The parameter estimates are taken as the subsample median value.

\subsubsection{Phase Transition Algorithm}

The application of an LPPL model to a system relies on an analogy between the process and critical phase transitions.
An explicit algorithm was developed in \cite{VanBov98} that conditions the data with an exponential detrending step, motivated by data processing in condensed matter applications.
The detrending step was also used in \cite{ZhoSor04}.

Let $s(t)$ be the observed time series.
Let $\hat{A},\hat{B},\hat{m}$ be best estimates for parameter values in an exponential trend, written as
\begin{equation}
	s(t) = A + B\exp(-m t)
\end{equation}
Let $r(t)$ be the model residuals equal to $s(t)-\hat{s}(t)$.
These are fit to a reduced LPPL model that assumes $m \rightarrow 0$ and a mean of zero, valid for residual processes.
The \textit{log-divergent} form of the LPPL has the following functional form
\begin{equation}
 r(t) = B \ln(t- t_c)[1+ D\cos(\omega ln(t- t_c)+\psi)]
\end{equation}
\\
\textbf{Phase Transition Algorithm:}
\begin{equation*}
	\{ x_1 , \dots, x_n \} \rightarrow \{ \hat{A}, \hat{B}, \hat{m}\} \rightarrow \{ \hat{B}, \hat{t}_{c},\hat{\omega},\hat{\psi},\hat{C}\}
\end{equation*}

The model is extendable to varying window-length samples, where best fits are again chosen as the median parameter value.

\section{ABCDE Model}
\label{sec:ABCDE}

\begin{figure}[h!]
\centering
	(a)\\
\includegraphics[scale=0.7]{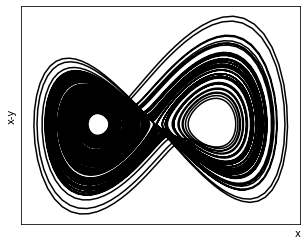}\\
	(b)\\
\includegraphics[scale=0.7]{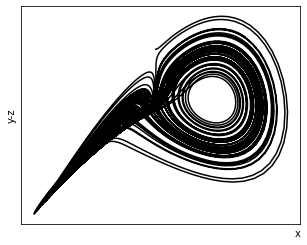}\\
	(c)\\
\includegraphics[scale=0.7]{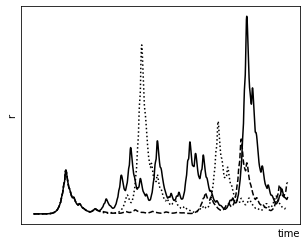}
	{\caption{{\bf ABCDE Exhibits Strange Attractor.} \\
	(a) Projection of Lorenz subsystem in ABCDE model relevant for simulational values near the phase transition.
	Initial parameters: $(x,y,z,r,\theta)=(0,1,2,1,5.03999)$.
	(b) Projection onto $x,y-z$ axis shows $x=\text{const.},y-z=\text{const.}$ as a recurrent point.
	(c) Three time series simulations in $r$ show intermittent deviations from $r=0$\label{strangeattractor}}}
\end{figure}

We discuss a chaotic system which exhibits self-organized behavior and phase transitions.
The ABCDE model \cite{Ken1976, FriHak92, Die2007} is a five-variable chaotic system originally proposed to study the dynamo effect.
A Lorenz subsystem is present within the ABCDE model, along with a dissipative subsystem, and the two are additively coupled.
We present a simplified version that has only one-way dependence between the Lorenz and dissipative subsystems.
The process exhibits intense intermittency when the control parameter, $\epsilon$, nears a transitional point.
For the certain parameter choices we will use in our simulations \cite{Die2007}, the transition occurs at $\epsilon=5$.\\

\textbf{ABCDE Model:}
\begin{eqnarray*}
	\dot{x} &=& \sigma (-x + y)\\
	\dot{y} &=& -y + (\rho - z)x\\
	\dot{z} &=& -\beta z + xy\\
	\dot{b}_1 &=& -\epsilon a_1 b_1 + \alpha x b_2\\
	\dot{b}_2 &=& -\epsilon a_2 b_2 + \alpha x b_1\\
\end{eqnarray*}
  
We introduce the change of variables:
\begin{eqnarray*}
	b_1 &=& rcosh(\theta)\\
	b_2 &=& rsinh(\theta)
\end{eqnarray*}

The new equations for the dissipative subsystem have the expression \cite{Die2007}:
\begin{eqnarray*}
	\dot{r}&=& \epsilon r [ -a_1+(a_2 - a_1)sinh^2(\theta)]\\
	\dot{\psi}&=& -\epsilon(a_2 -a_1) sinh(\theta) cosh(\theta) +\alpha x
\end{eqnarray*}

This is a five-variable model of deterministic chaos in the system variables $\{x,y,z,r,\theta\}$ parameterized by $\{ \sigma, \rho,\beta,a_1,a_2,\alpha, \epsilon\}$.

\subsection{Non-linear Phase Transitions}
\label{sec:phase}

Fixing $\sigma=10,\rho=2.667, \beta=28, a_1=0.1,a_2=0.2$, and $\epsilon=4.94$, near, but below that transitonal value $\epsilon=5$.
Fig. \ref{strangeattractor}a exhibits that the strange attractor present in the Lorenz subsystem is present and relevant for the chosen parameter values.

Figures \ref{strangeattractor}b,c provide simulational evidence for intermittency in $r$.  The graphical projection onto the $x, y-z$ axis shows that the point $x=\text{const.},y-z=\text{const.}$ is recurrent. 
This region is correlated with small values of $r$, so that $r$ exhibits intermittent and chaotic deviations from a neighborhood about zero.

\section{Methodology}
\label{sec:methodology}

We formulate our V\&V methodology around a precise statement of the LPPL estimation problem.
In repeated samples of $r(t)$ under varying initial conditions, the goal is to forecast the critical time $t_c$ under preceeding estimation windows that range in distance from the critical event.
Our definition of a singularity is drawn from applications of the LPPL model to financial markets, but it can be applicably rephrased for applications to geophysical processes.
A \textbf{critical event} is the start of a $-15\%$ drawdown.
The \textbf{peak value} is the value at the critical event.

Our statistical procedure follows repeated simulations of the ABCDE model:\\
\begin{enumerate}[topsep=0pt,itemsep=0ex,partopsep=1ex,parsep=1ex]
	\item Save time series of $r$
	\item Record start of the last drawdown as $t_c$
	\item Denote end of second to last drawdown as start of analysis
	\item End analysis window when $r$ achieves half, one-third, or one-quarter (commisserate with Feigenbaum's criticism) of peak value
	\item For subsamples of the analysis window:
		\begin{enumerate}
			\item Forecast $\hat{t}_c$ with subordination and phase transition algorithms
			\item Record median parameter estimates
		\end{enumerate}
	\item Compare aggregate mean absolute forecast errors with a standard t-test
\end{enumerate}

\subsection{Feigenbaum's Criticism: The Relevant Sample Window}

\cite{Fei2001} found that LPPL model fits were not particular to periods near a critical event.
But were well fit some distance away from the transition point.
The criticism is well articulated, and questions ex-post analysis set near a known transition.

Commiserate with this criticism, we compare three sample windows set increasingly distant from the simulated transition, rejecting the criticism if pairwise corrected $t$-tests reveal similar out-of-sample forecast accuracy.

\section{Results and Discussion}
\label{sec:results}

Our first result is categorically rejecting the phase transition algorithm.
An exponential detrend is relevant on some datasets, but produces errors in our simulations that are two orders of magnitude larger than the subordinated algorithm.

We restrict our attention to testing the Feigenbaum criticism.

We accept the Feigenbaum criticism if $\hat{t}_c$ accuracy deteriorates in distance to the critical event.
We test the hypothesis by fitting median sample fits across three windows: two distant from the transition, but near-enough to be comparable (25\% and 33\% peak value) and the last abutting the singular event (50\% peak value).

If we cannot reject the equality of sample errors between the three samples than we find that the LPPL model has zeroed in on a process dynamic prevalent throughout the sample period, and not an artefact in the near-vicinity of critical events.
This would constituite a rejection of Feigenbaum's criticism.

\begin{center}
\begin{tabular}{|l|l|l|l|}
	\multicolumn{1}{c}{\bfseries Hypothesis} & \multicolumn{1}{c}{\bfseries P-value} & \multicolumn{1}{c}{\bfseries P-value*} & \multicolumn{1}{c}{ \bfseries N}  \\ \hline
	$|\hat{t}_c - t_c|_{50\%} = |\hat{t}_c - t_c|_{33\%}$ & 0.49 & 0.98 & 565\\ \hline
	$|\hat{t}_c-t_c|_{50\%} = |\hat{t}_c-t_c|_{25\%}$ & 0.35 & >1 &  - \\ \hline
	$|\hat{t}_c - t_c|_{25\%} = |\hat{t}_c - t_c|_{33\%}$ & 0.81 & 0.81 &  - \\ \hline
\end{tabular}
\end{center}

P-values for multiple t-tests are corrected by the Holm-Bonferroni method \cite{Hol1979}.
$N$ is the sample size and number of simulations, which is equal across tests.
We fail to reject the null hypotheses that mean absolute sample errors for $\hat{t}_c$ across different sample windows are pairwise equal.

\section{Conclusion}
\label{sec:conclusion}

In concluding, we stress the importance of subsample re-estimation for the LPPL model, corroborating \cite{BreChaPei13, MarketCrash, SorCau15}.
The latter develops a confidence score based on subsample parameter variability.
These tools are broadly applicable to statistical learning in high-dimensional settings.

We record minor, but not statistically significant, decreases in mean absolute estimation error for $\hat{t}_c$ as the sample window approaches the critical transition.

The critical points exhibited in our choice of ABCDE parametrization are extreme when compared to financial crashes.
The recurrence of small values of $r$ means intermittent total disaster from an investor's perspective.
Further investigation may develop a parametrization more directly comparable to prior work.

\cite{VanBov98} and \cite{MarketCrash} are examples of papers that develop heuristics to qualify LPPL best fits for applied data analysis, where process noise, competing dynamics, or unmodelable structures complicate model interpretation. 
Our approach is extendable to the open problem of validating LPPL-derived signals.

\bibliographystyle{siam}
\bibliography{refs}

\end{document}